\documentclass[12pt]{article}
\usepackage{amsmath,amsfonts,amssymb,graphicx}
\newcommand {\be}{\begin{equation}}
\newcommand {\ee}{\end {equation}}
\begin{document}

\hspace{9 cm}{\bf NTUW 09.20 }\\

\begin{center}
\vspace {1 cm}
\LARGE{ \bf {A Simple Quark Model for $\gamma p \rightarrow K^+ \Lambda^0$ }}\\
\vspace{.25 in}
{E. M. Henley* and Kevin Pham }\\
{\small Department of Physics and Institute for Nuclear Theory, Box 351560,\\
 University of Washington, Seattle, WA 98195-1560 }\\
\end{center}
\vspace{0.25 in}
\begin{abstract}
{The photoproduction of the $\Lambda^0 $and $K^+$ from circularly polarized photons on protons is discussed in a simple quark model; we compare the results to experiments.}
\end{abstract}

\section{Introduction}           

There has been considerable interest in the photoproduction of strangeness on the proton. Both $\Sigma$ and $\Lambda$ baryons are produced. Here we will focus on $\Lambda^0$ production. What it is hoped to learn is the mechanism for strangeness creation. We present a simple quark model to try to understand the experimental results. In the next section. we briefly describe the experiments that have been carried out. In Section 3, we present some of the theoretical models for the $\Lambda$ and kaon photoproduction. In Section 4 we present our simple quark model and in Section 5 we give our results      and compare them to experiment. In Section 6 we present a non-relativistic approximation for the process we are considering, and in Section 7, we give our conclusions.  

\section {Experiments}

Experiments have been  carried out in Japan (LEPS)\cite{LEPS}, ELSA at Bonn 
(SAPHIR)\cite{SAPH}, and Jefferson Lab (CLAS)\cite{CLAS}. Some of the data is 
with linearly polarized photons\cite{SAPH} and some with circularly polarized 
gammas \cite {CLAS}. The first CLAS results were differential cross section measurements for $- 0.85 \leq cos \theta_K(c.m) \leq 0.95$ \cite{CLAS}. Later, they used incident  circularly polarized photons of c.m. energies from  1.6 to 2.53 GeV, or lab energies $0.9 < E_\gamma < 2.94$ GeV.  The circular polarization of the photons is maximal at the bremsstrahlung end point and falls slowly with decreasing $E_\gamma$, with $0.440 < P_\circ <0.995$. The CLAS experimenters measured both polarization transfer along the direction of the photon and perpendicular to it. The polarization of the $\Lambda$ was measured via the asymmetry in the decay $\Lambda^0 \rightarrow p \pi^-$.

The  c.m. frame chosen by the CLAS group \cite{CLAS} is shown in Fig.1. We will also use it. The c.m. angular distribution, the energy dependence of the induced polarization ($P_y$), the polarization transfers in the x- and z-directions, $C_x$, and $C_z$ were measured as a function of c.m. angle of kaon emission.

\begin {figure}
\begin{center}
\includegraphics[width=2.5 in]{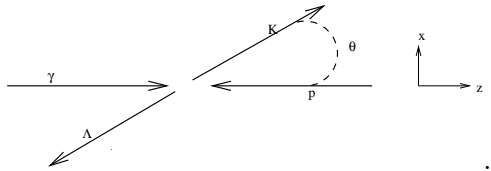}.
\end{center}
\caption{cm frame of reference}
\label{figure 1}
\end{figure}

One of the features of the experimental results that caught our attention was that the measurement of $C_z$ gave approximately $ 100 \% $ in the direction of the photon \cite{CLAS}

\be
 \frac {d\sigma}{d\Omega_K} = \frac {d\sigma}{d\Omega}_K |_{unpol} [1 +\sigma_y P + P_\circ (C_x \sigma_x +C_z \sigma_z)]\; ,
\ee
where
$\sigma_x, \sigma_y,$ and $\sigma_z $ are components of the Pauli spin matrix along the axes illustarted in Fig. 1, and $P_\circ$ is the circular polarization of the incident photon. 
There are some discrepancies between SAPHIR and CLAS  data (see \cite{LEPS}), but we will not dwell on this feature. 

\section {Theoretical analyses}

Most theoretical papers  use effective Lagrangian models with many baryon resonances (spin 1/2, 3/2, 5/2) in both, or either, $s$ (e.g., N*, $\Delta*$) and $u$ (e.g.,Y*)channels  \cite { CLAS}--see Fig. 2-- to fit the data. The authors hope to learn resonance information from fitting the data. However, as pointed out, the resonance parameters depend on the method of analysis (see e.g.,\cite{SIR,M&S}). Since resonance coupling constants and the widths of most decay channels  to $1/2^+$ are not well known, there are quite a few adjustable  parameters. Mart and Bennhold \cite{M&B} use four baryon resonances. The $D_{13}(1895)$ is predicted to have a particularly strong coupling to the $K^+ \Lambda^0$. Mart and Sulaksono \cite{M&S} use many more resonances, as do de la Puente, Maxwell, and Raue \cite{PMR}. Hadronic form factors, cut-off masses, and means to satisfy gauge invariance need to be determined. Ireland, Janssesn and Ryckebusch  discuss an  algorithm for the analysis of N* resonances in the photoproduction \cite{IJR}. They also need the N* resonance at 1.9 GeV. Janssen et al. \cite{J} discuss the role of hyperon resonances.  Chiang et al.  \cite {CSTL} use dynamical coupled channel models,(e.g., $\gamma N \rightarrow \pi N \rightarrow KN)$ as do  D\'{i}az, Saghai,  Lee, and Tabakin \cite{DS}, and Shklyar, Lenske, and Mosel \cite{SLM}. Guidal, Laget, and Vanderhaeghen  use Reggeized t-channel exchanges (K and K*)\cite{GLV}. Some authors use a Reggeized resonance approach \cite {Ryck}. Shyam, Scholten, and Lenske \cite{SSL} use a coupled channel effective Lagrangian method with many coupled two-body final states.  Finally, Li \cite {L} uses a non-relativistic quark model with coupling to resonances to calculate kaon photoproduction. There are thus many approaches that have been tried, most of them with many parameters that can be adjusted.

\begin{figure}
\begin{center}
\includegraphics[width= 4 in] {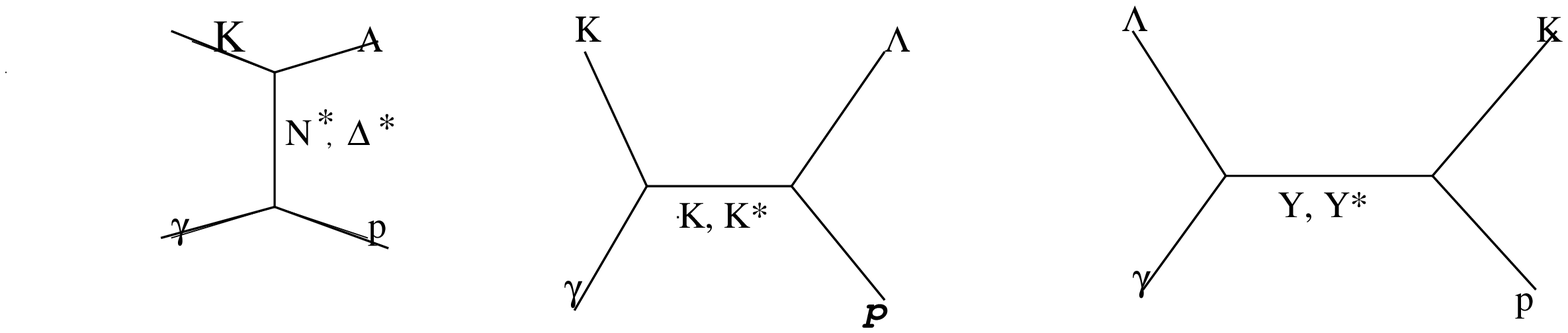}
\end{center}
\caption{resonance models used to explain the $\gamma p \rightarrow \Lambda^0 K^+$ data. Y indicates a strange baryon. }
\label{figure 2}
\end{figure}

\begin{figure}
\begin{center}
\includegraphics[width= 1.2 in] {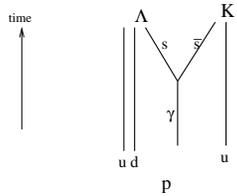}.
\end{center}
\caption{Lowest order quark model used in our calculation}
\label{figure 3}
\end{figure}

Few theorists consider quark models (perhaps because the energy is relatively low). An exception is Keiner, \cite{theor2} who carries out calculations up to $E_\gamma$ = 1.9 GeV, but does not consider polarization transfer, but only the differential and total cross sections. Alkofer et al \cite{ A} also use a quark model, but with intermediate N* resonances. A simpler quark  model came to our mind, because it suggests approximately 100\% $C_z$.

\section{Model}

In our model the $ud$ and $u$ quarks are spectators and only the $s$ and $\bar s$ play a direct role. The reason for concentrating on the $\Lambda$ as a test of the model now becomes clear. It is only the $ud$ in the spin and isospin zero channel in the proton which contributes to the reaction $ \gamma p \rightarrow K^+ \Lambda^0$ \cite{Alb}. The spin-isospin 1 combination contributes to the $\Sigma$ , but not to the $\Lambda$. Thus, the spin of the $\Lambda$ is carried totally by the strange quark. The $s$ and $\bar{ s}$ share the spin of the photon in this picture; since the coupling of the photon is directly to the $s$ and $\bar{s}$, this suggests that $ C_z \sim  1 $. 

Of course, it is not possible to conserve both energy and momentum with the lowest order diagram, Fig. 3, but the quarks are bound, and this binding helps to conserve both energy and momentum. To lowest order, we thus expect $C_x=0$ and $C_z\ \sim 1$. Calculations  bear this out. In addition to the basic diagram, we  have included rescattering corrections in order to obtain a non-vanishing $C_x$. There are several possible rescattering diagrams, as shown in Figs. 4 and 5. Neither of the diagrams of Fig. 4 involve the $s$ quark; thus one expects the $\Lambda$ to remain approximately 100\% polarized in the z-direction and $C_x$ =0. Calculations agree with this expectation. Rescattering diagrams which affect the $s$ quark are shown in Fig. 5. For simplicity we take the rescattering to be due to the exchange of a scalar boson rather than a gluon exchange  between the quarks, e.g.between  $s$ quark and the $ud $ diquark, the $s$ and $u$ quarks, or the $s$ and $\bar{s}$. We parameterize the rescattering by a Gaussian in momentum space, $ V= g^2 exp(-t^2/\alpha^2)$, where $t$ is the momentum of the exchanged scalar, and $\alpha$ is a parameter.  

\begin{figure}
\begin{center}
\includegraphics[width= 2.0 in]{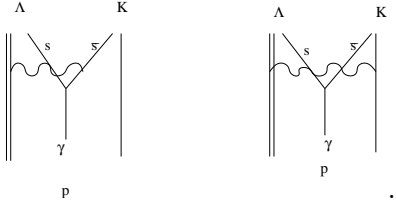}.
\end{center}
\caption{Some rescattering diagrams that do not involve s quarks}
\label{figure 4}
\end{figure}

\begin{figure}
\begin{center}
\includegraphics[width= 2.5 in] {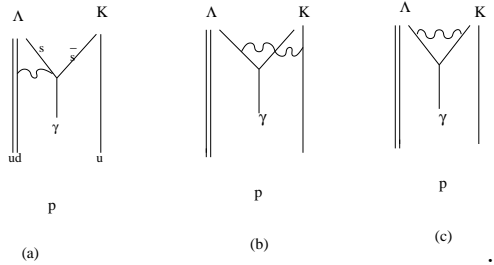}.
\end{center}
\caption{Rescattering diagrams that involve the $s$ quark}
\label{figure 5}
\end{figure}

We have carried out calculations for both Figs. 5a  and 5b. When the differences are small, we present the results only for Fig. 5a. The gamma algebra for the Feynman diagrams of Fig. 5a and 5b  are carried out with the help of Mathematica. It is interesting that, in our simple quark model $P_y$ remains zero despite the rescattering corrections. This is probably due to our choice of a scalar, rather than a vector gluon, for the rescattering exchange. We have neglected final state interactions between the $K^+$ and $\Lambda^0$. 

In order to minimize the number of parameters, we have taken the mass of the strange quark and diquark to be the same, m= 0.35 GeV, and that of the u quark to be  1/2 of this mass. The binding (or distribution function) of the quarks is parameterized by a Gaussian of width  $\alpha^2 =(.55 )^2 GeV^{-2}$, which we also take as the width of the Gaussian in the rescattering. . 

For the root diagram, we need to obtain the trace
\be
Tr \mathcal{M}^\dagger(1+\gamma_5 \not{P}) \mathcal M = Tr(\frac{\not s +m}{2m} \not \epsilon (1 + \gamma_5 \not P) \not \epsilon \frac{-\not\bar{s}  +m}{2m} )\;  ,
\ee 
where $s $ and $\bar{s}$ are the momenta of the named quarks, and P is a polarization vector in any spatial direction; it turns out that, as expected, the only non-vanishing contribution is in the z-direction. This squared matrix element must still be integrated over the quark distribution function.

We take $s_x= p_\Lambda x sin\theta  $ and $s_z = p_\Lambda x cos\theta $, where $\theta$ is the scattering angle and $ 0 \leq x = x_F \leq 1$. This approximation may be somewhat questionable at the energies we are considering. We also take $\bar{s} = k -s$, and $-i(\epsilon_x \epsilon_y - \epsilon_ y \epsilon_x) = P_\circ$.   We use momentum conservation to obtain $ud = p_\Lambda - s, u= p_K - \bar{s} = p_p - ud$. The binding or quark distribution functions are given by $\phi$,  expressed totally in terms of $s$, the momentum of the $s$ quark  and the momenta of the $\gamma, p, K^+, and \Lambda^0$ momenta, by means of momentum conservation,
\be
\phi \propto exp(-(s -p_\Lambda)^2/\alpha^2) exp(-(\bar{s}-2p_K/3)^2/\alpha^2) exp(-(p_\Lambda - s -2p_p)^2/\alpha^2) .
\ee
The momenta in this proportionality are all 4-vectors.

For the case of Fig.5a we need the trace 
\be
Tr \mathcal{M}^\dagger (1+\gamma_5 \not{P}) \mathcal {M}=\\ 
Tr[g^2\not{\epsilon}\frac{\not{r} +m}{r^2-m^2}\frac{\not{s}+m}{2m}(1 + \gamma_5 \not{P})\frac{\not{r} + m}{r^2-m^2} \not{\epsilon}\frac{-\not\bar{s} +m}{2m}]\; ;
\ee
a similar trace is needed for Fig. 5b. Here $r$ is the momentum of the intermediate $s$ state. The distribution function for the second order diagram, Fig. 5a, is like that to first order, but $\bar{s} =k-r$ and $r  = s-t$, where $t$ is the momentum carried by the scalar, 
\be 
V(t) = g^2 exp(-t^2/\alpha^2) \; .
\ee
 For the intermediate state propagator, we used the pole approximation; that is we take the integration over the zero-component to give the value at the pole. 
\be
\int{\frac{dr_0}{r^2 - m^2}}=\int_{- \infty}^\infty dr_0 \frac{1}{r_0^2- \vec{r}^2 -m^2}= \frac{i \pi}{\sqrt{\vec{r}^2 + m^2}}
\ee

Keeping the full 4-dimensions in the numerical integration, one gets infinity due to the pole. 
Thus, we  presently have three parameters, the width of the Gaussian, the constant $g$,  and the mass of the strange quark. Of course, the choice of the up quark as having a mass of 1/2 that of the strange quark is another parameter. 

The differentila cross section is 
\be
\frac{M_p}{E_p}\frac{(2 \pi)^4}{2k}|\mathcal{M^\dagger M}|^2 \frac{d^3p_\Lambda}{(2 \pi)^3} \frac{d^3p_K}{2 \omega_K(2\pi)^3}\frac{M_\Lambda}{E_\Lambda}\delta^4(p_K + p_\Lambda - k -p_p),
\ee
where k is the momentum of the photon, $ p_p$ and $E_p$ refer to the proton, $M_\Lambda, E_\Lambda$ refer to the $\Lambda^0$, and $\omega_k$ is the energy of the kaon.
                              
\section {Results}
Our  results for typical energies are presented together with some of the CLAS data in a series of figures. The angular distributions at incident photon energies of 1.7 and 2.3 GeV are compared to experiment in Figs. 6 and 7, respectively. The fit is remarkably good for the CLAS data, but not so for the SAPHIR output. We do not get as large a rise in the backward direction of the differential cross section as the experimental results show. We have also done calculations at 2.0 and 2.6 GeV. The magnitude of the differential cross section $d \sigma/d(cos \theta) $ at $0^o$ and 1.7GeV is found to be $0.27 \mu$barns at first order, 1.9  $\mu$barns for case Fig.5(a) and 0.4 $\mu$barns for case Fig.5(b). Experimentally, $d\sigma/d(cos \theta)$ = 1.8 $\mu$barns at $cos \theta$=0.9. As seen from the figures, the fit is quite good for Fig. 5(a).  

\begin{figure}
\begin{center}
\includegraphics[width= 2.5 in] {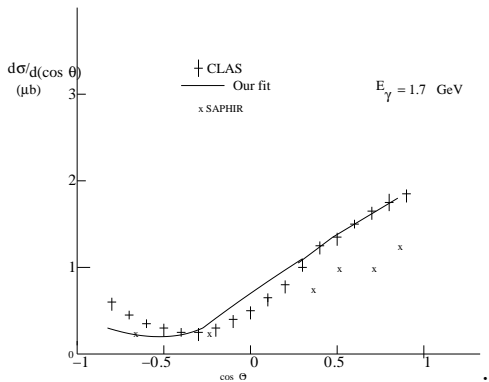}.
\end{center}
\caption{Differential cross section for $E_\gamma$ =1.7 GeV (W=2.02 GeV) and for the case of Fig. 5(a); the  curve is similar for Fig.5(b), except for the scale}
\label{figure 6}
\end{figure}

\begin{figure}
\begin{center}
\includegraphics[width= 2.5 in] {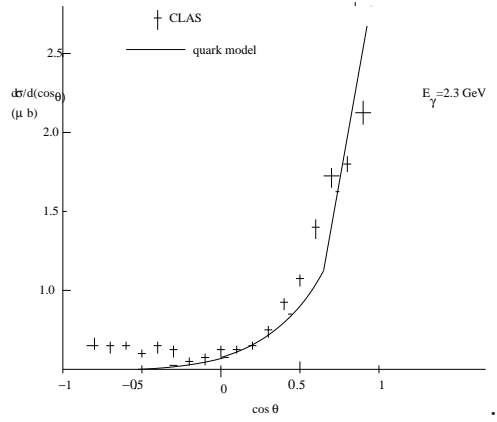}.
\end{center}
\caption{Differential cross section at W$\approx$ 2.3 GeV  ($E_\gamma \approx$2.3 GeV). The calculated differential cross section at $0^o$ is 2.8 $\mu b$ vs an experimental one of 1.6 $\mu b$ at $cos\theta =0.9$.}
\label{figure 7 }
\end{figure}

In Figs. 8 and 9, we show the fits to the transferred polarization in the direction of the photon at 1.7 and 2.3 GeV photon energies. Both lowest order and rescattering corrections are presented. To lowest order, $C_z$ increases as the scattering angle increases, whereas with rescattering it decreases and changes sign at large angles. In both cases $|C_z| $ gets to be larger than unity at backward angles. This shows that there are problems with our model. 

\begin{figure}
\begin{center}
\includegraphics[width=4 in] {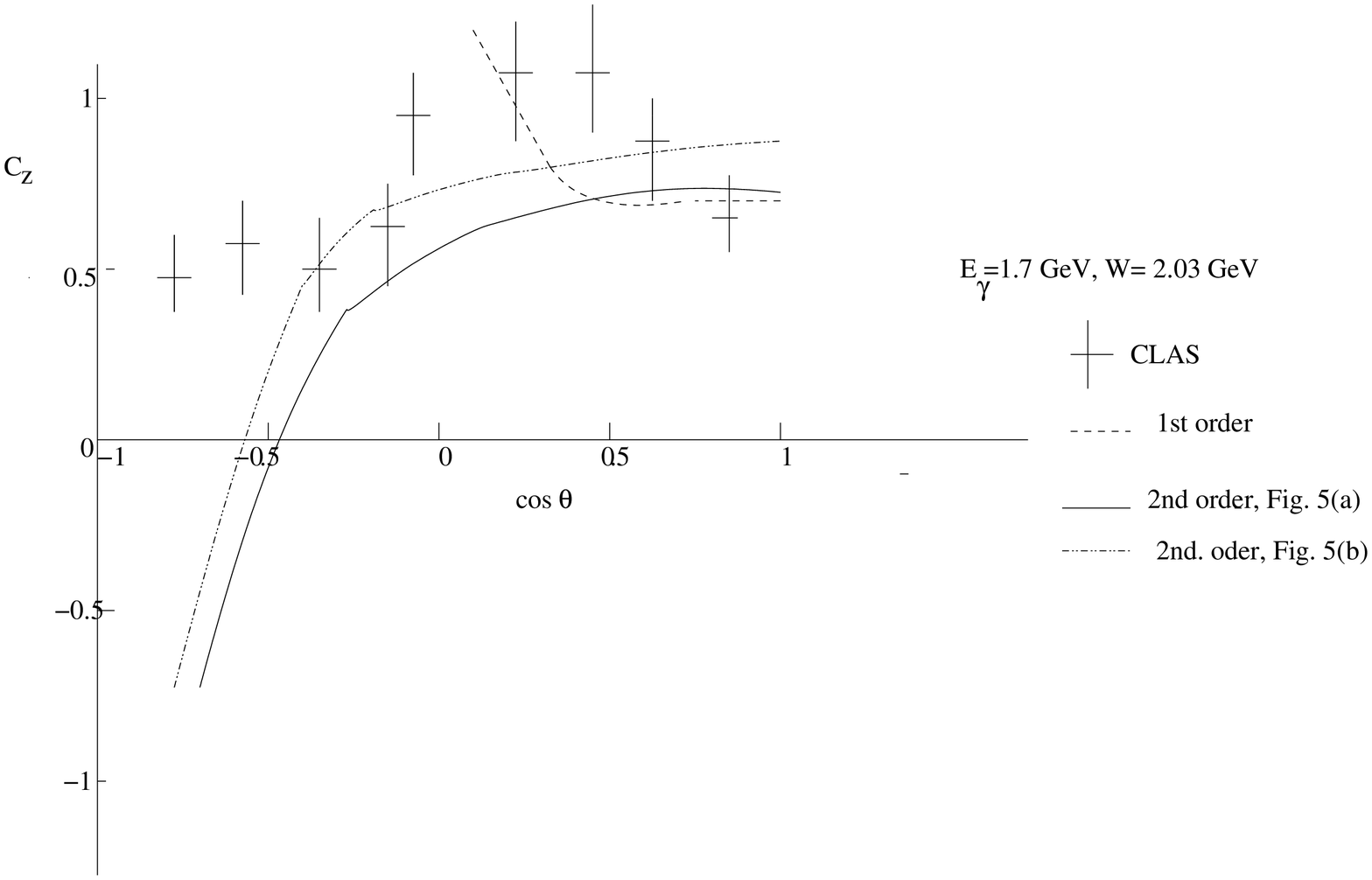}.
\end{center}
\caption{$C_z$ at W=2.03 GeV, ($E_\gamma \approx$ 1.7 GeV). Cases a and b refer to Fig.5}
\label{figure8}
\end{figure}

\begin{figure}
\begin{center}
\includegraphics[width= 4 in] {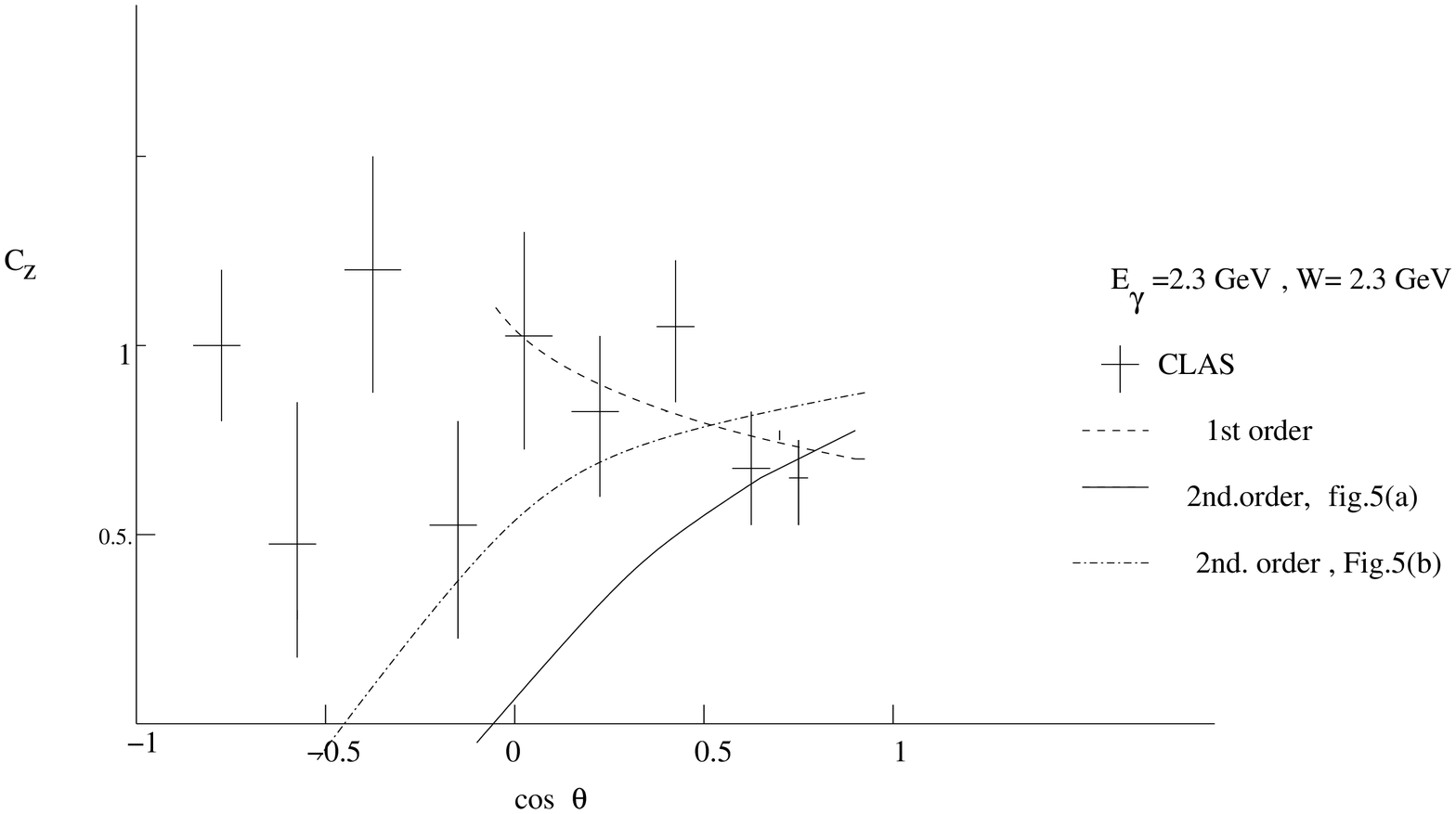}.
\end{center}
\caption{$C_z$ at W $\approx$ 2.3 GeV, ($E_\gamma$ $\approx$ 2.3 GeV). Cases a and b refer to Fig.5}
\label{figure9}
\end{figure}

The polarization transfer in the x-direction (in the scattering plane) is shown in Fig.10 for $E_\gamma$ = 2 GeV (W=2.17 GeV); it is compared to experiment. The theoretical value of $C_x$ vanishes at $0^o$(as it should, since x is not defined at $0^o$ and$ 180^o$), and grows in magnitude with angle. Like $C_z$ it gets to be larger than 1 in magnitude at back angles. The experimental data fluctuates widely.

\begin{figure}
\begin{center}
\includegraphics[width= 4 in] {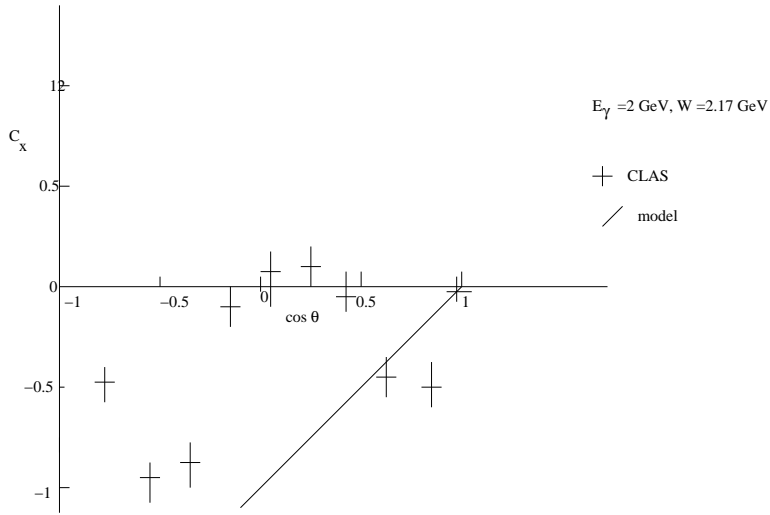}.
\end{center}
\caption{$C_x$ at W $\approx$ 2.17 GeV ($E_\gamma \approx 2.0 GeV$)}            \label{figure 10}
\end{figure}
\newpage

\section{Non-relativistic Approximation}

We have also carried out the calculation for our model in lowest order in a non-relativistic approximation and present the results in this section. 

Non-relativistically, the current which couples to the photon is taken as
\be  
\vec{j} = \frac{\vec{s}-\vec{\bar{s}}}{2m} + i \frac{\vec{\sigma} \times \vec{k}}{2m}\; ,
\ee
where $\vec{s}$ and $\vec{\bar{s}}$ are the momenta of the $s$ and $\bar{s}$ quarks, $ \vec{k}$ is the photon momentum, $\vec{\sigma}$ is the Pauli spin operator, and m is the mass of the strange quark. In the non-relativistic case, the trace 
\be
\mathcal{M}^2=Tr(\vec{\epsilon}^*\cdot\vec{j}^* \frac{1 + \vec{\sigma} \cdot \vec{P}}{2} \vec{\epsilon}\cdot \vec{j})  .
\ee 
is needed when we omit the polarization. 
\be
\mathcal{M}^2=\frac{ s_x^2  -2 s_x \bar{s}_x + \bar{s}_x^2}{8m^2} \; ;
\ee
here $s_x$, $\bar{s}_x$ are the x-components of the 3-momenta of the s and $\bar{s}$ quarks. 

The trace vanishes for $P_x$ and $P_y$,  so that there is no polarization transfer in the x-direction nor is there an induced polarization in the y-direction. The polarization transfer in the z-direction arises from the square of the last term in Eq. (7); the cross term does not contribute. 
 
The trace must be integrated over the quark distribution function, which is like that in Eq.(3), but in only 3-dimensions. For the polarization in the z-direction we obtain simply $P_z = k^2/(8m^2)$. $C_z$ is equal to $P_z$ divided by $ |\mathcal{M}|^2$. We do not show the results because they exceed unity at all angles. The arbitrarily normalized differential cross section is similar to that of the relativistic model, but rises more in the backward direction. 

\section{Conclusions}

We presented a simple quark model to explain the photoproduction of $K^+$ and $\Lambda^0$ on protons. It is the only straightforward quark model of this reaction for incient polarized photons, as far as we know. Most other calculations use many resonances to fit the data.  The differential cross-section  and polarization transfer are given and compared to some of the experimental data. Despite some remaining problems, the fits we obtain to the CLAS data are reasonable, considering the simplicity of our model..

Finally, a brief summary of a non-relativistic approximation is shown. The shortcomings of this approximation are presented. 

To improve our model, we need to use gluon exchange between quarks instead of a scalar.

\vspace {0.2 in}

$^*$Supported, in part, by the Department of Energy

\end{document}